\documentclass[runningheads]{llncs}
\usepackage[T1]{fontenc}

\usepackage{graphicx,verbatim}

\usepackage{amsmath}  % Now load amsmath safely
\usepackage{amsfonts}
\usepackage{amssymb}
\usepackage{graphicx}
\usepackage{dsfont}

\usepackage{graphicx}
\usepackage{subcaption}
\usepackage{orcidlink}
\usepackage{booktabs}
\usepackage{enumitem}

\begin{document}
\title{Log-Euclidean Frameworks for Smooth Brain Connectivity Trajectories}

\author{Olivier Bisson\orcidlink{0009-0007-1727-7048} \and
Yanis Aeschlimann\orcidlink{0009-0009-8427-1409} \and
Samuel Deslauriers-Gauthier\orcidlink{0000-0003-2781-121X} \and
Xavier Pennec\orcidlink{0000-0002-6617-7664}}
\authorrunning{O. Bisson et al.}
\institute{Université Côte d’Azur, Inria, France}
\maketitle

\begin{abstract}
The brain is often studied from a network perspective, where functional activity is assessed using functional Magnetic Resonance Imaging (fMRI) to estimate connectivity between predefined neuronal regions. Functional connectivity can be represented by correlation matrices computed over time, where each matrix captures the Pearson correlation between the mean fMRI signals of different regions within a sliding window. We introduce several Log-Euclidean Riemannian framework for constructing smooth approximations of functional brain connectivity trajectories. Representing dynamic functional connectivity as time series of full-rank correlation matrices, we leverage recent theoretical Log-Euclidean diffeomorphisms to map these trajectories in practice into Euclidean spaces where polynomial regression becomes feasible. Pulling back the regressed curve ensures that each estimated point remains a valid correlation matrix, enabling a smooth, interpretable, and geometrically consistent approximation of the original brain connectivity dynamics. Experiments on fMRI-derived connectivity trajectories demonstrate the geometric consistency and computational efficiency of our approach.  

\end{abstract}

\section{Introduction}

The brain cortex has been intensively studied over the past decades from a graph perspective, where the nodes are homogeneous groups of neurons segregated into regions using atlases. Considering an atlas dividing the cortex into $N$ regions, the connectivity between the regions are stored in a $N \times N$ connectivity matrix, called a connectome~\cite{HumanConnectome}. Here, we focus on functional connectomes (FC), built from resting-state functional Magnetic Resonance Imaging (rs-fMRI), where the connectivity is the Pearson correlation of the averaged fMRI timeseries between each pair of regions. In many studies on functional connectivity at rest, the brain network of a subject is characterized by a unique FC, using the correlation computed on the whole fMRI timeseries. However, recent studies have described the underlying brain network as a temporal succession of multiple functional networks' activation~\cite{DynamicConnectomes}. In this context, we compute a series of correlation matrices using a sliding window approach with overlap, resulting in a time series of full-rank correlation matrices that reflect the evolving functional brain connectivity. While symmetric positive definite (SPD) matrices are commonly used in geometric data analysis, the limitations of the Euclidean metric have led to the development of alternative Riemannian metrics, such as the affine-invariant~\cite{pennec2006riemannian} and Log-Euclidean metrics~\cite{arsigny2006log}. Recent work has extended these geometric tools to full-rank correlation matrices, enabling the definition of diffeomorphic mappings into Euclidean vector spaces. In this paper, we leverage the Riemannian structures developed in \cite{Thanwerdas2024} to propose an \emph{intrinsic polynomial regression} of timeseries of correlation matrices. By performing polynomial regression in the image space of Log-Euclidean diffeomorphisms and pulling back the result, we construct smooth and interpretable trajectories of correlation matrices that approximate dynamic functional connectomes. Experiments show that Euclidean regression produces immediately non-physical negative eigenvalues and that rescaling SPD matrices to satisfy the unit-diagonal constraint of full-rank correlation matrices can alter the correlation coefficients by up to $7.51\%$, whereas our method avoids reprojection.

\section{HCP Dataset, Data Processing}
\begin{figure}[h!]
    \centering
    \includegraphics[width=0.75\textwidth]{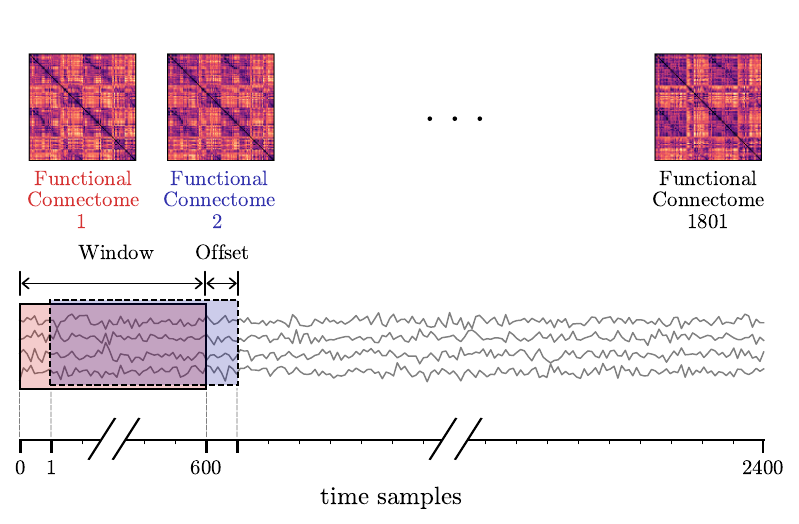}
\caption{Illustration of dynamic functional connectome generation computed from resting-state fMRI signals using sliding windows.}
\label{fig:connectome_generation}
\end{figure}

We selected the first $100$ subjects from the Human Connectome Project (HCP) S1200 release. The imaging protocols for rs-fMRI have been described in detail in~\cite{VANESSEN20122222}.
The atlas of $7$ networks and $100$ regions proposed by~\cite{schaefer2018local} is used to segment the cortical regions. The pipeline of connectome generation is as follows.
Rs-fMRI data has been preprocessed by the HCP Minimal Preprocessing Pipeline~\cite{HCP:minimali} of the HCP. The timeseries of left-right and right-left phase encoding are concatenated along the time axis, leading to a total of $2400$ time samples. Irrelevant parts of the fMRI signal are removed with a bandpass filter with cut-off frequencies of $0.01$ and $0.1 \mathrm{Hz}$. The resulting fMRI signals are averaged over the regions of the atlas. The correlation matrices are computed on processed fMRI signals using Pearson’s coefficient on truncated timeseries using sliding windows of $600$ time samples and with an offset of $1$, resulting in $T=1801$ FCs per subject. The values for the sliding windows have been experimentally chosen to keep the eigenvalues of the FCs between $10^2$ and $10^{-1}$. See Figure~\ref{fig:connectome_generation} for an overview of the construction of the connectomes.

\section{Riemannian Metrics on Full-Rank Correlation Matrices}

Let $\{C_t\}_{t=1}^{T}\subset\mathrm{Cor}^+(n)$ denote a time series of dynamic functional connectomes obtained from sliding-window resting-state fMRI. The central problem is to compute \emph{smooth} approximations of this trajectory while \emph{staying} inside the manifold of full-rank correlation matrices. Two flat diffeomorphic parametrisations make this possible:
\begin{itemize}[leftmargin=2.1em,itemsep=0.3em]
\item the \emph{off-log} map  
      $\mathrm{Log}\colon\mathrm{Cor}^+(n)\!\longrightarrow\!\mathrm{Hol}(n)$, obtained by zeroing the diagonal after taking the matrix logarithm;
\item the \emph{log-scaling} map  
      $\mathrm{Log}^{\bullet}\colon\mathrm{Cor}^+(n)\!\longrightarrow\!\mathrm{Row}_0(n)$, obtained by optimal diagonal scaling prior to the logarithm.
\end{itemize}
Throughout the paper we adopt the following matrix–space notation:

\setlength{\tabcolsep}{8pt}         % horizontal padding
\renewcommand{\arraystretch}{1.2}   % vertical padding
\begin{center}
\begin{tabular}{@{}ll@{}}
\toprule
Symbol             & Matrix space \\ 
\midrule
$\mathrm{Cor}^{+}$ & smooth manifold of full-rank correlation matrices \\
$\mathrm{Hol}$     & vector space of hollow matrices (symmetric, zero diagonal) \\
$\mathrm{Row}_{0}$ & vector space of symmetric matrices with zero row-sum \\
$\mathcal{S}$      & vector space of symmetric matrices \\
$\mathcal{S}^{+}$  & smooth manifold of symmetric positive-definite matrices \\
\bottomrule
\end{tabular}
\end{center}

\subsection{The Off-Log Diffeomorphism}

It was proven in~\cite{Thanwerdas2024} that the off-log bijection defined in~\cite{Archakov2020} $\mathrm{Log} : \text{Cor}^+(n) \longrightarrow \text{Hol}(n)$ between the vector space of hollow matrices and full-rank correlation matrices is a smooth diffeomorphism. Since the off-log diffeomorphism is equivariant under permutations, it enables to pullback families of permutation-invariant inner products on hollow matrices to full-rank correlation matrices. 

\begin{theorem}\cite{Thanwerdas2024}
For $n \geq 4$, permutation-invariant inner products on $\mathrm{Hol}(n)$ are defined by quadratic forms for $X \in \mathrm{Hol}(n)$:
\begin{align*}
q(X) = \alpha \mathrm{tr}(X^2) + \beta \mathds{1}^\top(X^2)\mathds{1} + \gamma \mathds{1}^\top(X)^2\mathds{1}, 
\end{align*}
with $\alpha > 0$, $2 \alpha + (n-2) \beta > 0$ and $\alpha + (n-1)(\beta + n \gamma) >0$. For $n=3$ set $\alpha =0$ and for $n=2$ set $\alpha = \beta = 0$. 
\end{theorem}

Given a symmetric matrix $S \in \mathrm{Sym}(n)$, there exists a unique diagonal matrix $D \in \mathrm{Diag}(n)$ such that $\exp(D + S)$ is a full-rank correlation matrix, thereby defining a smooth parametrization of the space of full-rank correlation matrices.

\begin{theorem}\cite{Archakov2020}
For all $ S \in \mathrm{Sym}(n) $, there exists a unique $ D \in \mathrm{Diag}(n) $ such that $ \exp(D+S) \in \mathrm{Cor}^+(n) $. We denote this mapping as
$ %%$\begin{align*}
    \mathcal{D} : S \in \mathrm{Sym}(n) \longmapsto D \in \mathrm{Diag}(n).
$ %%\end{align*}
\end{theorem}

\subsubsection{Computation of $\mathcal{D}$} An inductive algorithm to compute $D$ that converges linearly was proposed in~\cite{Archakov2020}: write $D_0 = 0$ and define $D_{k+1} = D_k - \log (\mathrm{Diag}(\exp (D + S))) \in \mathrm{Diag}(n)$. The algorithm stops at step $k$ whenever $| D_{k} - D_{k-1} |_E \leq \epsilon$, where $| \cdot |_E$ is the usual Euclidean norm, and $\epsilon > 0$ is a fixed threshold for convergence. 

\subsubsection{Properties of the Off-Log Mapping} Let us now specify the diffeomorphism, its inverse and their differential given in~\cite{Thanwerdas2024}. For all $C \in \mathrm{Cor}^+(n)$ and $S, X, Y \in \mathrm{Hol}(n)$, define $H^0 \in \mathrm{Sym}^+(n)$ by $H^0_{il} = \sum_{j, k} P_{ij} P_{ik} P_{lj} P_{lk} \exp^{(1)}(\delta_j, \delta_k)$, where $P \in \mathcal{O}(n)$ and $\Delta = \mathrm{diag}(\delta_1, \ldots, \delta_n) \in \mathrm{Diag}(n)$ are such that $\mathcal{D}(S) + S = P\Delta P^\top$ and where $\exp^{(1)}$ denotes the first divided difference of the exponential.

\begin{itemize}
\item[]{\textbf{Diffeomorphism:}} $\mathrm{Log} : C \in \mathrm{Cor}^+(n) \longmapsto \mathrm{Off} \circ \log (C) \in   \mathrm{Hol}(n) $.
\item[]{\textbf{Inverse diffeomorphism:}} \\$\mathrm{Exp} = \pi_{\vert \mathrm{Hol}(n)} :  S \in \mathrm{Hol}(n) \longmapsto \exp (\mathcal{D}(S) + S) \in \mathrm{Cor}^+(n)$.
\item[]{\textbf{Tangent diffeomorphism:}} \\$d_C \mathrm{Log} : X \in \mathrm{Hol}(n) \longmapsto \mathrm{Off} \circ d_C \log (X) \in \mathrm{Hol}(n)$.
\item[]{\textbf{Inverse tangent diffeomorphism:}} \\$d_S \mathrm{Exp} : Y \in \mathrm{Hol}(n) \longmapsto d_{\mathcal{D}(S) + S} \exp (Y + d_S \mathcal{D}(Y)) \in \mathrm{Hol}(n)$,
\end{itemize}
where $d_S \mathcal{D}(Y)$ is given by $ -\mathrm{diag}\bigl((H^0)^{-1} \mathrm{Diag}(d_{\mathcal{D}(S) + S} \exp(Y)) \mathds{1}\bigr)$. 

\subsubsection{Off-Log Metrics} The family of off-log metrics on full-rank correlation matrices are obtained as the pullback of a Riemannian metric characterized by a permutation-invariant quadratic form on hollow matrices.

\begin{definition}\cite{Thanwerdas2024}
An off-log metric on $\mathrm{Cor}^+(n)$ is the pullback metric of a permutation-invariant inner product characterized by a quadratic form $q$ on $\mathrm{Hol}(n)$ by the diffeomorphism $\mathrm{Log} : \mathrm{Cor}^+(n) \longrightarrow \mathrm{Hol}(n)$.
\end{definition}

\begin{theorem}\cite{Thanwerdas2024}
Let $g$ be an off-log metric on $\mathrm{Cor}^+(n)$ characterized by a permutation-invariant quadratic form $q$ on $\mathrm{Hol}(n)$. The Riemannian operations associated to $g$ are, for all $C, C', C_i \in \text{Cor}^+(n)$, $X \in \mathrm{Hol}(n)$, $t \in \mathbb{R}$:
\begin{enumerate}
\item[]{\textbf{\emph{Inner product: }}} $g_C(X, X) = q(d_C \mathrm{Log}(X))$.
\item[]{\textbf{\emph{Exponential map: }}} $\mathrm{Exp}_C(X) = \mathrm{Exp}(\mathrm{Log} (C) + d_C (\mathrm{Log} (X)))$.
\item[]{\textbf{\emph{Logarithm map: }}} $\mathrm{Log}_C(C') = d_{\mathrm{Log}(C)} \mathrm{Exp}(\mathrm{Log}(C') - \mathrm{Log} (C))$.
\item[]{\textbf{\emph{Geodesic: }}} $\gamma_{C \rightarrow C'}(t) = \mathrm{Exp}((1-t)\mathrm{Log}(C) + t\mathrm{Log}(C'))$.
\item[]{\textbf{\emph{Squared distance: }}} $d(C, C')^2 = q(\mathrm{Log}(C') - \mathrm{Log}(C))$.
\item[]{\textbf{\emph{Parallel transport: }}} $\Pi_{C \rightarrow C'}(X) = (d_{C'} \mathrm{Log})^{-1}(d_C \mathrm{Log}(X))$.
\item[]{\textbf{\emph{Curvature: }}} $R= 0$.
\item[]{\textbf{\emph{Fréchet mean: }}} $\bar{C}=\mathrm{Exp}(\frac{1}{n} \sum_{i=1}^{n} \mathrm{Log}(C_i))$.
\end{enumerate}
\end{theorem}

\subsection{The Log-Scaling Diffeomorphism}

The log-scaling diffeomorphism $\mathrm{Log}^* : \mathrm{Cor}^+(n) \to \mathrm{Row}_0(n)$, mapping full-rank correlation matrices to the space of symmetric matrices with null row sum, was first made explicit in \cite{Thanwerdas2024} building on the earlier works \cite{marshall1968} and \cite{johnson009}. Being equivariant under permutations, it enables the pullback of permutation-invariant inner products from $\mathrm{Row}_0(n)$ to $\mathrm{Cor}^+(n)$.

\begin{theorem}\cite{Thanwerdas2024}
For $n \geq 4$, permutation-invariant inner products on $\mathrm{Row}_0(n)$ are defined by quadratic forms for $Y \in \mathrm{Row}_0(n)$:
\begin{align*}
q^*(Y) = \alpha \mathrm{tr}(Y^2) + \beta \mathrm{tr}(\mathrm{Diag}(Y)^2) + \gamma \mathrm{tr}(Y)^2, 
\end{align*}
with $\alpha > 0$, $n \alpha + (n-2) \beta > 0$ and $n \alpha + (n-1)(\beta + n \gamma) >0$. For $n=3$ set $\alpha =0$ and for $n=2$ set $\alpha = \beta = 0$. 
\end{theorem}

Given an SPD matrix $\Sigma \in \mathrm{Sym}^+(n)$, there exists a unique positive diagonal matrix $\Delta \in \mathrm{Diag}^+(n)$ such that $\log(\Delta \Sigma \Delta)$ is a null row sum symmetric matrix, thereby defining a smooth parametrization of the space of full-rank correlation matrices via the inverse of $\mathrm{Log}^*$.

\begin{theorem}\cite{marshall1968}\cite{johnson009}
For all $\Sigma \in \mathrm{Sym}^+(n)$, there exists a unique $\Delta \in \mathrm{Diag}^+(n)$ such that $\log(\Delta \Sigma \Delta) \in \mathrm{Row}_0(n)$. We denote this mapping as 
\begin{align*}
    \mathcal{D}^* : \Sigma \in \mathrm{Sym}^+(n) \longmapsto \Delta \in \mathrm{Diag}^+(n)
\end{align*}
\end{theorem}

\subsubsection{Computation of $\mathcal{D}^*$} The numerical computation of $\mathcal{D}^*(\Sigma)$ for $\Sigma \in \mathrm{Sym}^+(n)$ is a well-known problem in the literature, commonly referred to as scaling a matrix to prescribed row sums. As observed in \cite{Thanwerdas2024}, this problem can be reformulated as a strictly convex optimization problem. Specifically, for a given $\Sigma \in \mathrm{Sym}^+(n)$, $D = \mathcal{D}^*(\Sigma)$ if and only if it minimizes the strictly convex function $F: D \in \mathrm{Diag}^+(n) \longmapsto \frac{1}{2} \mathds{1}^\top D^\top \Sigma D \mathds{1} - \mathrm{tr}(\log (D)) \in \mathbb{R}$. The gradient at $D \in \mathrm{Diag}^+(n)$ is $\nabla_D F = \Sigma D \mathds{1} - D^{-1} \mathds{1}$, and the Hessian is $H_D F = \Sigma + D^{-2}$. Given the closed-form expression for the Hessian, Newton's method provides an efficient numerical approach to computing $\mathcal{D}^*$.

\subsubsection{Properties of the Log-Scaling Mapping} Let us now specify the diffeomorphism, its inverse and their differential given in \cite{Thanwerdas2024}. For all $C \in \mathrm{Cor}^+(n)$, $X \in \mathrm{Hol}(n)$ and $S,  Y \in \mathrm{Row}_0(n)$ such that $\Sigma = \mathcal{D}^*(C) C \mathcal{D}^*(C) = \exp (S)$.

\begin{itemize}
\item[]{\textbf{Diffeomorphism:}} \\$\mathrm{Log}^* : C \in \mathrm{Cor}^+(n) \longmapsto \log (\mathcal{D}^*(C) C \mathcal{D}^*(C)) \in   \mathrm{Row}_0(n) $.
\item[]{\textbf{Inverse diffeomorphism:}} \\$\mathrm{Exp}^* : S \in \mathrm{Row}_0(n) \longmapsto \mathrm{Cor} \circ \exp (S) \in \mathrm{Cor}^+(n)$.
\item[]{\textbf{Tangent diffeomorphism:}} \\$d_C \mathrm{Log}^* : X \in \mathrm{Hol}(n) \longmapsto d_\Sigma \log (\Delta X \Delta + \frac{1}{2} (X^0 \Sigma + \Sigma X^0) ) \in \mathrm{Row}_0(n)$.
\item[]{\textbf{Inverse tangent diffeomorphism:}} $d_S \mathrm{Exp}^* : Y \in \mathrm{Row}_0(n) \longmapsto$\\
$ \Delta^{-1}\left( d_S\exp (Y) -\frac{1}{2}(\Delta^{-2} \mathrm{Diag}(d_S\exp (Y))\Sigma + \Sigma \mathrm{Diag}(d_S\exp (Y)) \Delta^{-2}) \right) \Delta^{-1} \in \mathrm{Hol}(n)$.
\end{itemize}
Where $\Delta = \mathrm{Diag}(\Sigma)^{1/2}$ and $X^0= -2 \mathrm{diag}\bigl((I_n + \Sigma)^{-1} \Delta X \Delta \mathds{1}\bigr)$. 

\subsubsection{Log-Scaled Metrics} The family of log-scaled metrics on full-rank correlation matrices are obtained as the pullback of a Riemannian metric characterized by a permutation-invariant quadratic form on symmetric null row sum matrices.

\begin{definition}\cite{Thanwerdas2024}
A log-scaled metric on $\mathrm{Cor}^+(n)$ is the pullback metric of a permutation-invariant inner product characterized by a quadratic form $q^*$ on $\mathrm{Row}_0(n)$ by the diffeomorphism $\mathrm{Log}^* : \mathrm{Cor}^+(n) \longrightarrow \mathrm{Row}_0(n)$.
\end{definition}

\begin{theorem}\cite{Thanwerdas2024}
Let $g^*$ be a log-scaled metric on $\emph{Cor}^+(n)$ characterized by a permutation-invariant quadratic form $q^*$ on $\emph{Row}_0(n)$. The Riemannian operations associated to $g^*$ are, for all $C, C', C_i \in \mathrm{Cor}^+(n)$, $X \in \mathrm{Hol}(n)$, $t \in \mathbb{R}$:
\begin{enumerate}
\item[]{\textbf{\emph{Inner product: }}} $g^*_C(X, X) = q^*(d_C \mathrm{Log}^*(X))$.
\item[]{\textbf{\emph{Exponential map: }}} $\mathrm{Exp}_C(X) = \mathrm{Exp}^*(\mathrm{Log}^* (C) + d_C (\mathrm{Log}^* (X)))$.
\item[]{\textbf{\emph{Logarithm map: }}} $\mathrm{Log}_C(C') = d_{\mathrm{Log}^*(C)} \mathrm{Exp}^*(\mathrm{Log}^*(C') - \mathrm{Log}^* (C))$.
\item[]{\textbf{\emph{Geodesic: }}} $\gamma_{C \rightarrow C'}(t) = \mathrm{Exp}^*((1-t)\mathrm{Log}^*(C) + t\mathrm{Log}^*(C'))$.
\item[]{\textbf{\emph{Squared distance: }}} $d(C, C')^2 = q(\mathrm{Log}^*(C') - \mathrm{Log}^*(C))$.
\item[]{\textbf{\emph{Parallel transport: }}} $\Pi_{C \rightarrow C'}(X) = (d_{C'} \mathrm{Log}^*)^{-1}(d_C \mathrm{Log}^*(X))$.
\item[]{\textbf{\emph{Curvature: }}} $R= 0$.
\item[]{\textbf{\emph{Fréchet mean: }}} $\bar{C}=\mathrm{Exp}^*(\frac{1}{n} \sum_{i=1}^{n} \mathrm{Log}^*(C_i))$.
\end{enumerate}
\end{theorem}

\section{Smooth Approximation of Brain Connectivity Trajectories Using Polynomial Regression}

\subsubsection{Methodology}  
We compare time-parameterized curves of correlation matrices obtained from rs-fMRI data before and after applying diffeomorphisms that induce off-log and log-scaling Riemannian metrics on full-rank correlation matrices, as well as the matrix logarithm, which induces a Log-Euclidean metric on SPD matrices. To this end, we transform the original curve using the corresponding diffeomorphism and apply a $3$D PCA to each transformed curve for visualization.

We also contribute an efficient implementation of these diffeomorphisms within the Python package \textit{geomstats}~\cite{JMLR:v21:19-027}. Applying the off-log diffeomorphism to a time series of $1801$ correlation matrices of size $100 \times 100$ takes approximately $1.24$ seconds, while its inverse takes $76.75$ seconds. The log-scaling diffeomorphism takes approximately $15.05$ seconds, and its inverse takes $1.26$ seconds.

\begin{figure}[h!]
    \centering
    \includegraphics[width=0.85\textwidth]{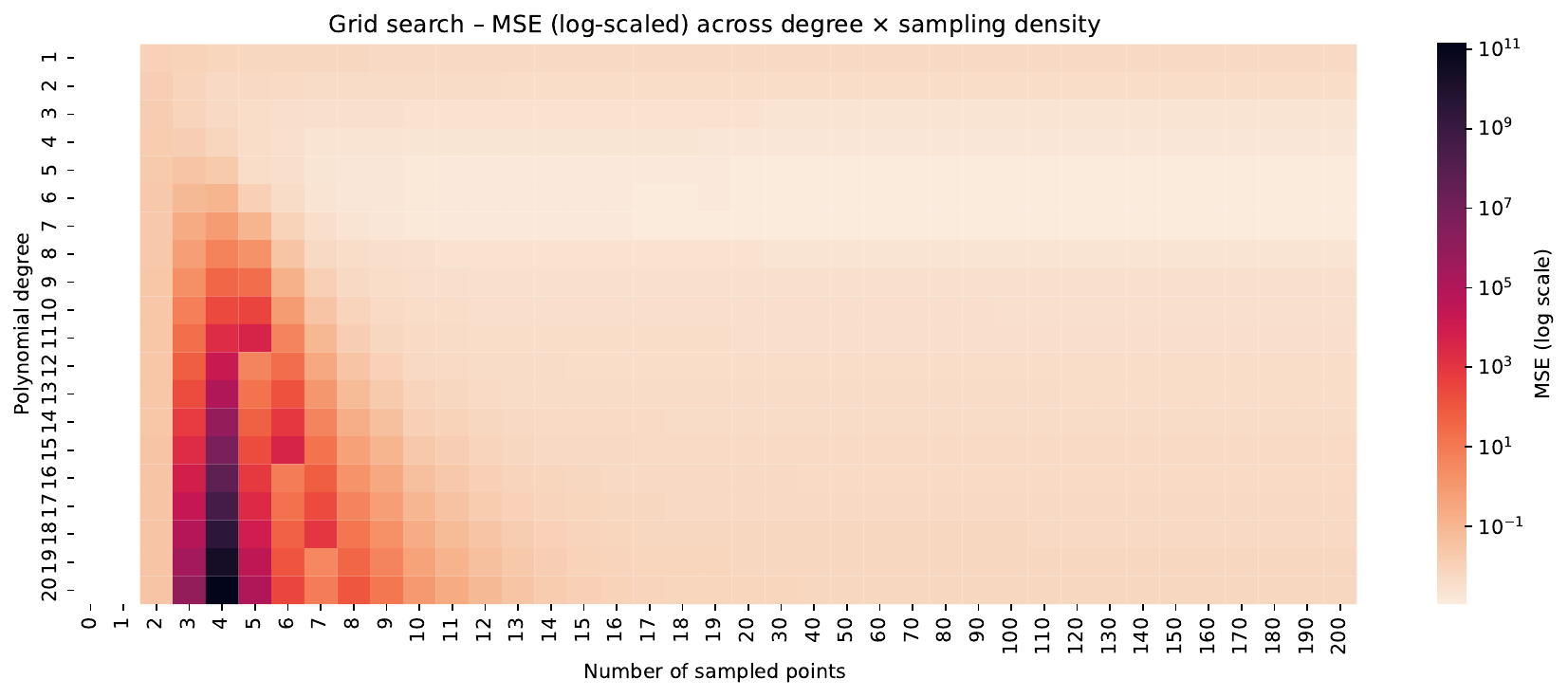}
\caption{Grid-search heat map of log-scaled mean-squared error (MSE) over polynomial degree (vertical axis) and sampling density (horizontal axis). Lighter cells correspond to lower error, revealing a broad minimum around degree $6$ once roughly $10$ or more sample points are used.}
\label{fig:grid_search}
\end{figure}

We perform polynomial regression of degree $6$ on $10$ evenly spaced time samples for both the original and transformed signals (Fig.~\ref{fig:pca_comparison}). The hyper-parameters (the polynomial degree and the number of samples) were chosen via a grid search that minimizes mean squared error (Fig.~\ref{fig:grid_search}). Finally, we pull back the regressed curve through the inverse diffeomorphism, yielding a smooth approximation of the original signal that stays within the manifold of full-rank correlation matrices.

In contrast, the SPD Log-Euclidean framework only provides a smooth approximation within the space of SPD matrices, which must be rescaled to be interpreted as connectomes, but this operation may modify the correlations by up to $7.51\%$ as shown in Fig.~\ref{fig:poly_pullback_scaling_results}. In the Euclidean framework, however, the regressed signal immediately exits the space of SPD matrices with negative eigenvalues—and, by extension, the space of full-rank correlation matrices—making true projection back into the space impossible.

\subsubsection{Visual Comparison of Curves}  
\begin{figure}[!b]
    \centering
    \vspace*{-3mm}
    \begin{subfigure}[b]{0.49\textwidth}
        \centering
        \includegraphics[width=\textwidth]{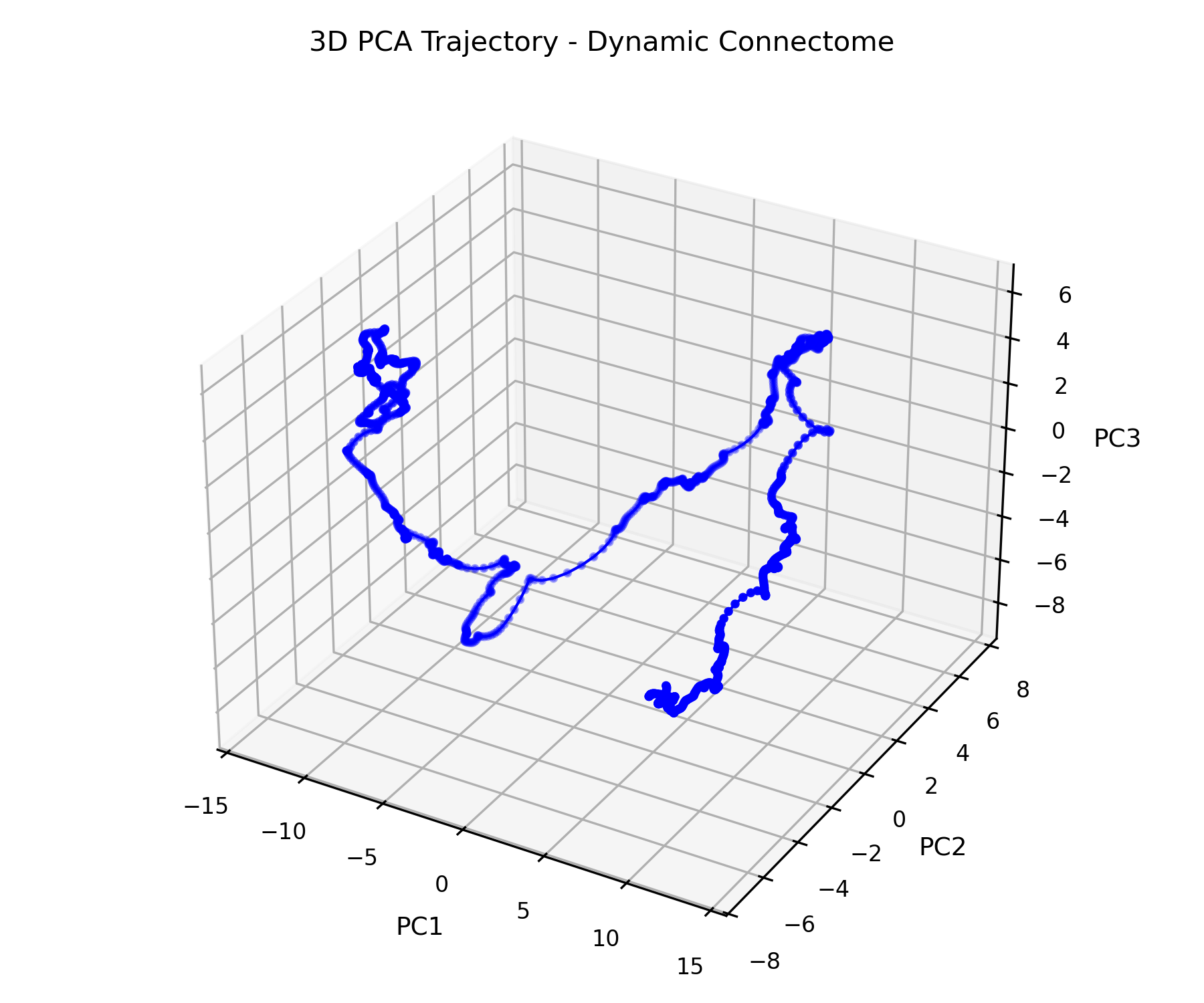}
        \caption{Original (correlation mat.)}
        \label{fig:all_c}
    \end{subfigure}
    \hfill
    \begin{subfigure}[b]{0.49\textwidth}
        \centering
        \includegraphics[width=\textwidth]{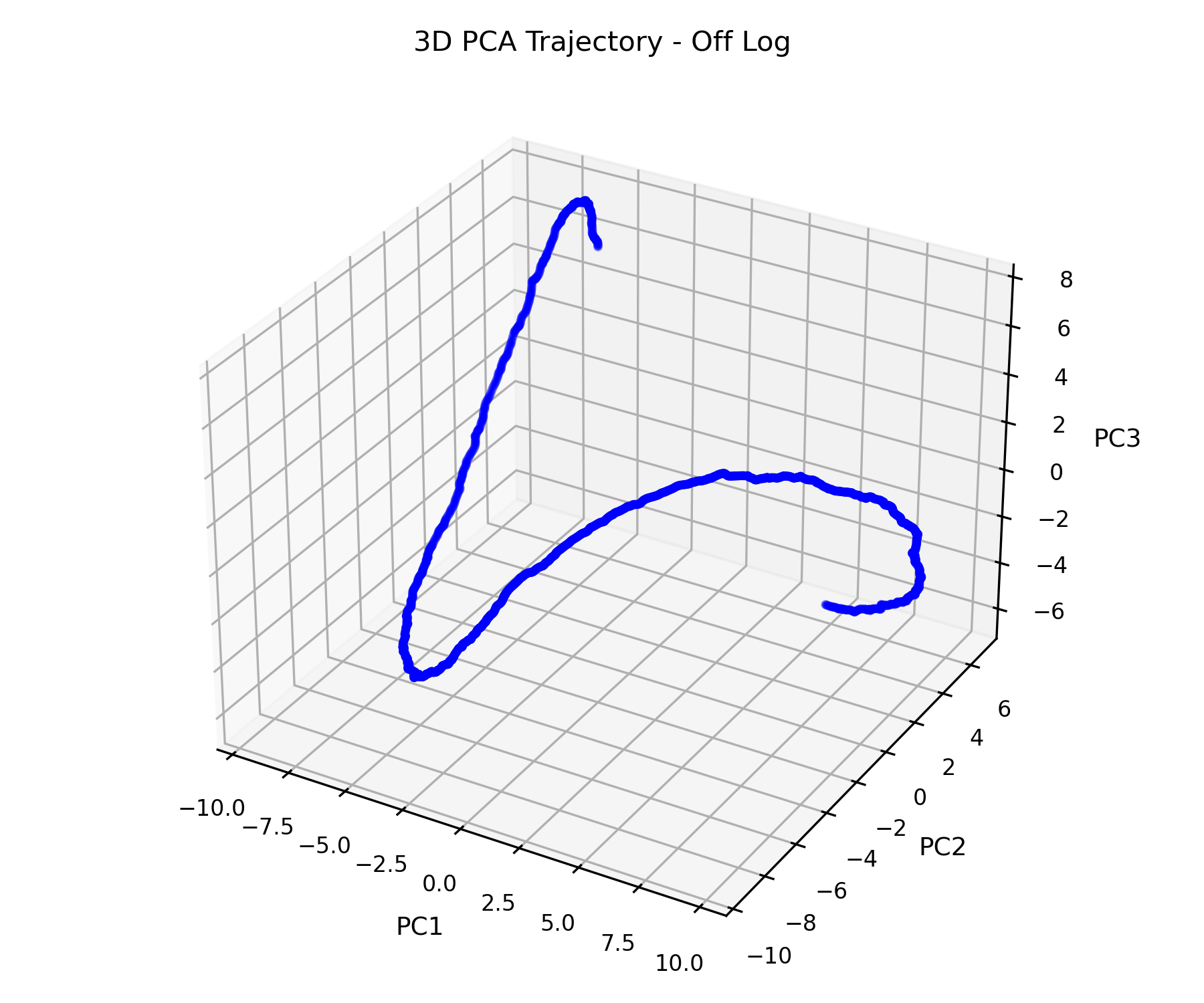}
        \caption{Off Log (hollow mat.)}
        \label{fig:off_log}
    \end{subfigure}
    \begin{subfigure}[b]{0.49\textwidth}
        \centering
        \includegraphics[width=\textwidth]{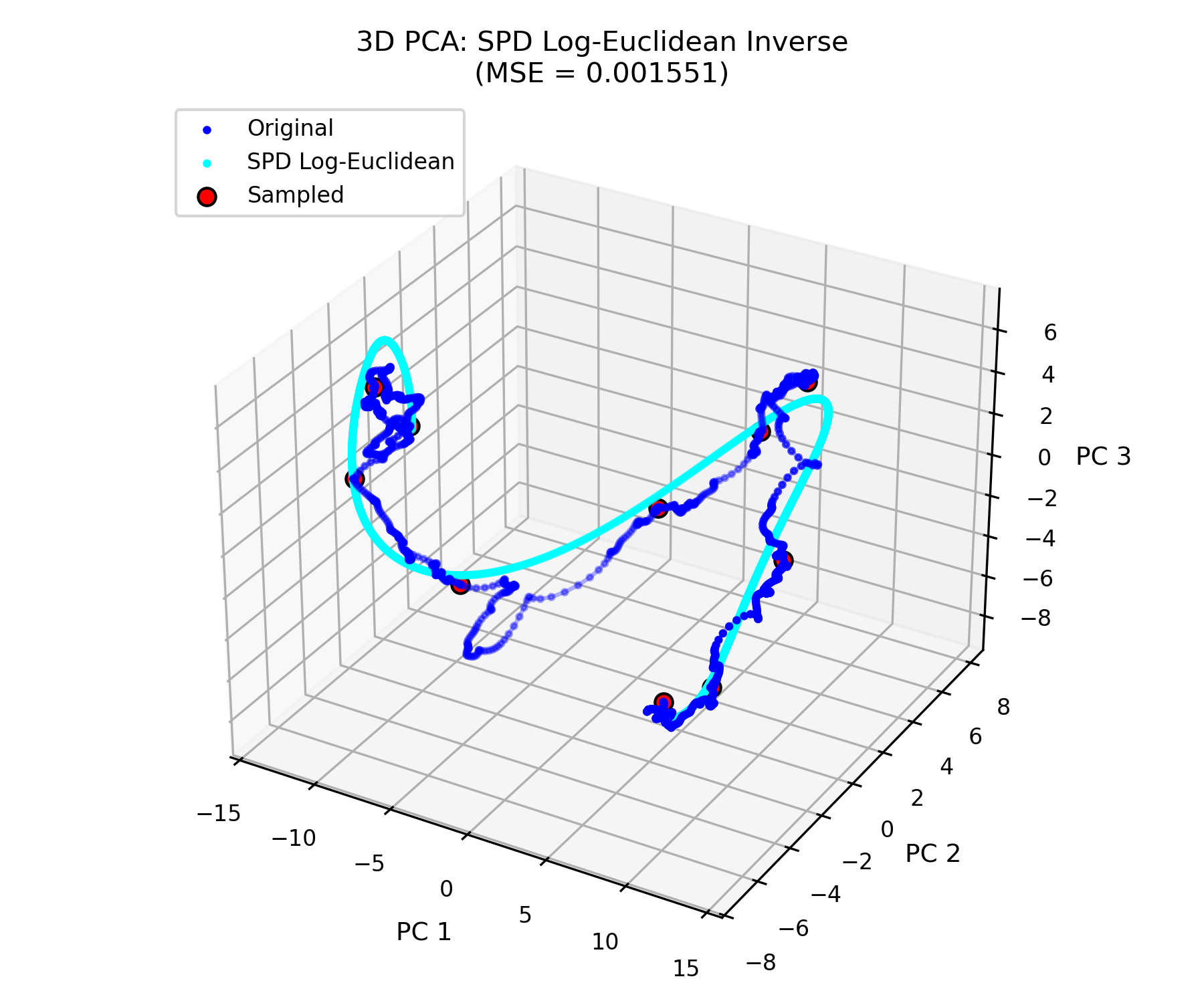}
        \caption{Log-Euclidean regression (SPD mat.)}
        \label{fig:poly_spd_original}
    \end{subfigure}
    \hfill
    \begin{subfigure}[b]{0.49\textwidth}
        \centering
        \includegraphics[width=\textwidth]{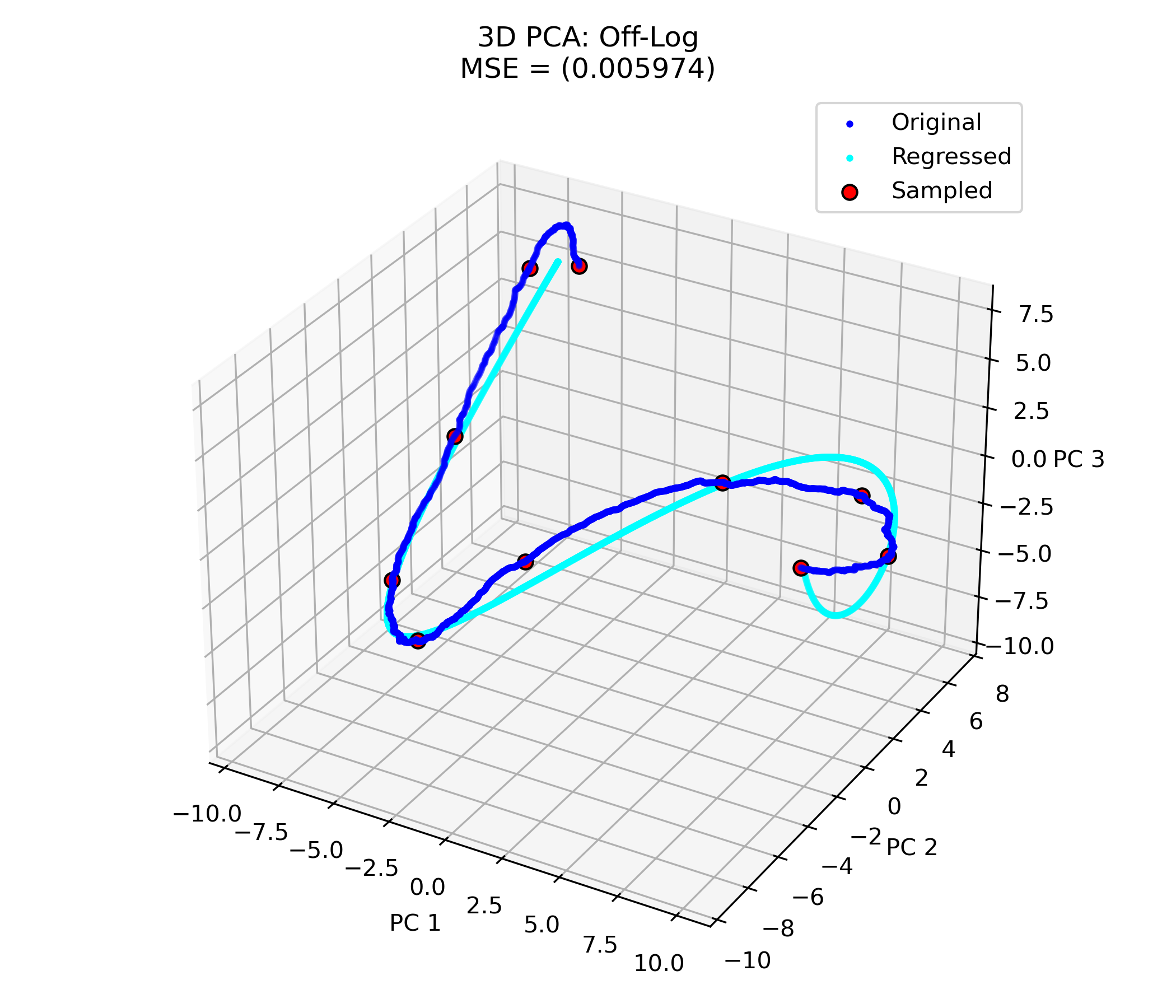}
        \caption{Off Log regression (hollow mat.)}
        \label{fig:poly_offlog}
    \end{subfigure}
    \caption{
    Left: Euclidean visualization; Right: Hollow (Off Log) visualization. Top: original data; Bottom: regressed curves. While the transformed trajectory (\ref{fig:off_log}) appears smoother, this does not reflect intrinsic temporal consistency. The regressed trajectory (\ref{fig:poly_offlog}) lies within the space of correlation matrices while the regressed matrices in (\ref{fig:poly_spd_original}) do not satisfy unit-diagonal constraint.
    }

    \label{fig:pca_comparison}
\end{figure}

Performing a $3$D PCA before and after transformation provides a 3D visualization of brain connectivity trajectories. A key observation is that, while the original signal appears continuous, it lacks regularity (i.e., differentiability); see Fig.~\ref{fig:pca_comparison}. In contrast, the transformed curves appear visually smoother, with points at time $t$ seemingly more evenly spaced relative to their neighbors at $t{-}1$ and $t{+}1$. However, this perceived regularity is largely an artifact of the PCA projection: pairwise Frobenius distance measurements reveal that the apparent spacing arises from the embedding in $\mathbb{R}^3$ and does not reflect true geometric distances between the matrices.

\begin{figure}[!t]
    \centering
    \begin{subfigure}[t]{0.75\textwidth}  %% {0.9\textwidth}
        %\centering
        %\includegraphics[width=0.75\textwidth, height=0.18\textheight]{min_eigenvalues_plot.png}
        \includegraphics[width=\textwidth]{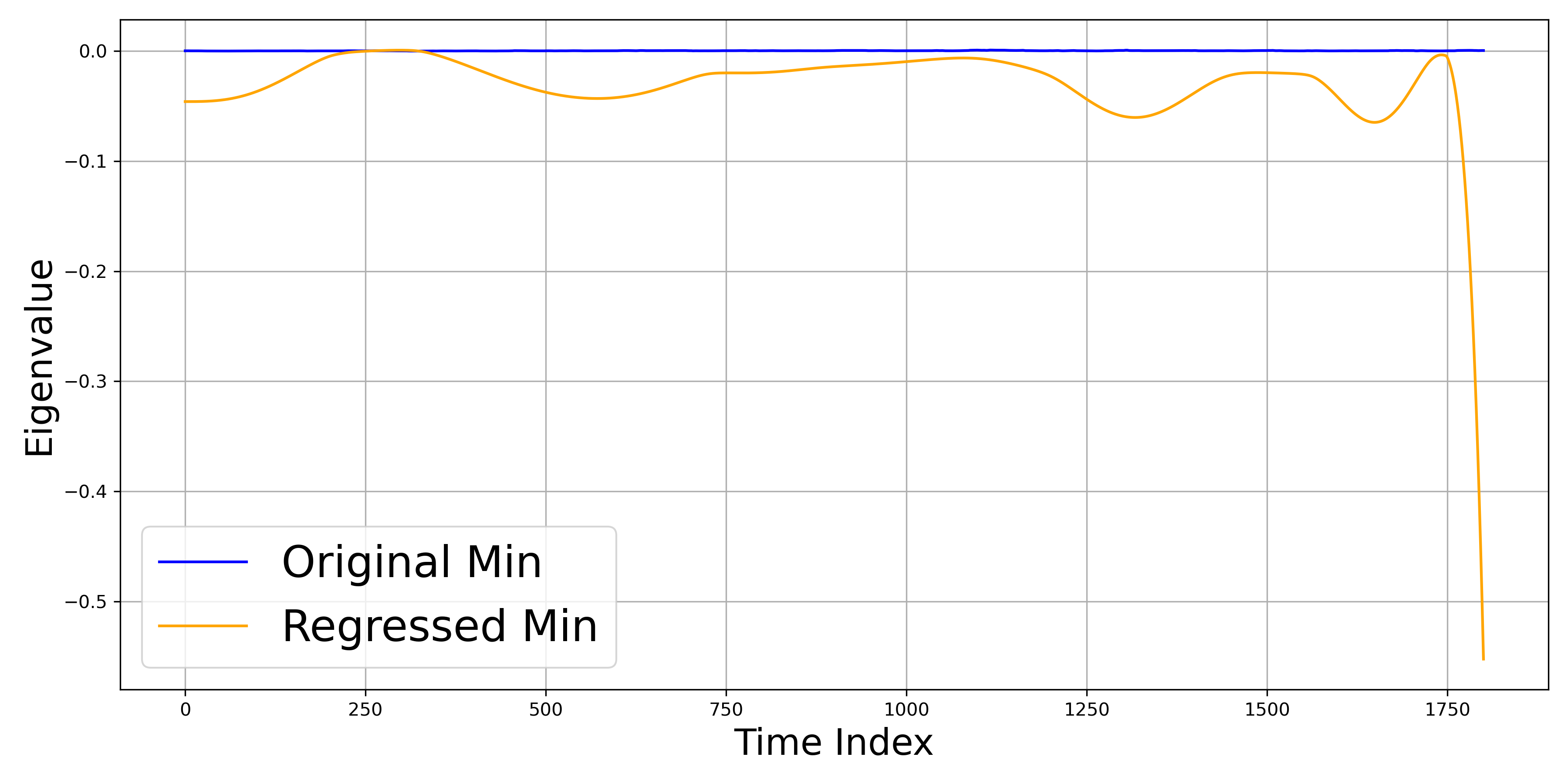}
        \caption{Minimum eigenvalues of original regressed curve over time}
        \label{fig:eigenvalue}
    \end{subfigure}
    \hfill
    %\vspace{0.7em}
    \begin{subfigure}[t]{0.75\textwidth}  %% {0.9\textwidth}
        %\centering
        %\includegraphics[width=0.75\textwidth]{scaling_factor_plot.png}
        \includegraphics[width=\textwidth]{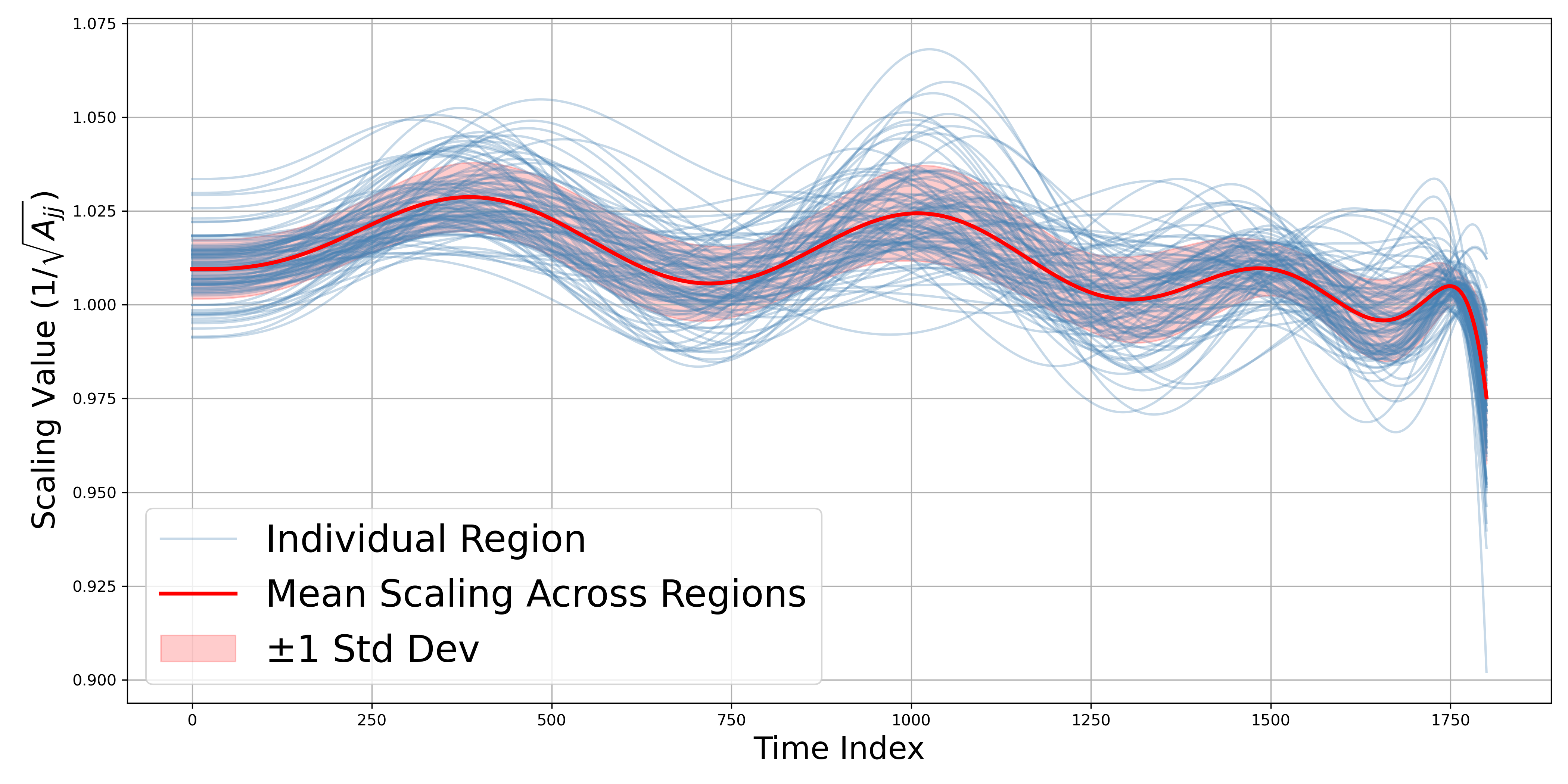}
        \caption{Scaling factors over time}
        \label{fig:scaling_factor}
    \end{subfigure}
    \caption{
    Polynomial regression of brain connectivity trajectories in Euclidean space violates positive definiteness, as shown by the minimum eigenvalues in \subref{fig:eigenvalue}. The SPD Log-Euclidean approach ensures positive definiteness, but projection back to correlation space alters the coefficients, as shown in \subref{fig:scaling_factor}, with relative deviation to the mean up to $7.51\%$.
    }
    \label{fig:poly_pullback_scaling_results}
\end{figure}

Even though the regressed curves show no apparent visual differences after inverse transformation, they are in fact clearly different. The regressed curve in the original space is mostly composed of indefinite symmetric matrices (see Fig.~\ref{fig:poly_pullback_scaling_results}), while the SPD-regressed curve requires reprojection to satisfy the unit-diagonal constraint of correlation matrices.

\enlargethispage{10mm}

\subsubsection{Polynomial Regression via Log-Euclidean Pullback}

As linear combinations of SPD matrices are not guaranteed to remain SPD or to satisfy the unit diagonal constraint required for full-rank correlation matrices we leverage the diffeomorphism $\varphi$ defining a Log-Euclidean-type Riemannian metric, which maps the manifold of correlation matrices into a Euclidean space. In this vector space, it becomes feasible to construct smooth polynomial regressions of the transformed trajectory. Specifically, given the time-series $\{C_t\}_{t=1}^{T}$ of correlation matrices, we compute $\{Z_t = \varphi(C_t)\}_{t=1}^{T}$ and fit a polynomial curve $t \mapsto P(t)$ in the Euclidean space via least-squares at ten evenly spaced time points. We then apply the inverse diffeomorphism $\varphi^{-1}$ to pull back the regressed curve to the original manifold: $\tilde{C}_t = \varphi^{-1}(P(t))$, yielding a smooth approximation of the original signal, where each regressed point $\tilde{C}_t$ is itself a valid correlation matrix. We obtain a smooth function of time whose values are guaranteed to lie in the space of full-rank correlation matrices, thus preserving interpretability at each time point.

\section{Conclusion and Perspectives}
We proposed an intrinsic method for polynomial regression of dynamic functional connectomes based on diffeomorphic log-Euclidean mappings of full-rank correlation matrices. Unlike approaches that project SPD matrices onto the correlation space post-regression, our method ensures that each regressed point remains a valid correlation matrix throughout. This provides a smooth and interpretable approximation of brain connectivity trajectories. Future work includes exploring alternative flat Riemannian metrics on full-rank correlation matrices, assessing their utility for neuroimaging applications, and benchmarking their computational costs. We also observed similar performance between the off-log and log-scaling metrics in our polynomial regressions; an in-depth study of the theoretical relationship between these two metrics is therefore necessary.

\paragraph{Fundings and Acknowledgments}
{\small
This work was supported by ERC grant \#786854 \textit{G-Statistics} from the European Research Council under the European Union’s Horizon 2020 research and innovation program, and by the French government through the \textit{3IA Côte d’Azur} Investments ANR-23-IACL-0001 managed by the National Research Agency (ANR). Data were provided by the Human Connectome Project, WU-Minn Consortium (Principal Investigators: David Van Essen and Kamil Ugurbil; 1U54MH0\break 91657) funded by the 16 NIH Institutes and Centers that support the NIH Blueprint for Neuroscience Research; and by the McDonnell Center for Systems Neuroscience at Washington University.}


\begin{thebibliography}{8}

\bibitem{Archakov2020}
Archakov, I., Hansen, P.R.: A new parametrization of correlation matrices. arXiv: Econometrics (2020).

\bibitem{Thanwerdas2024}
Thanwerdas, Y.: Permutation-invariant log-Euclidean geometries on full-rank correlation matrices. SIAM J. Matrix Anal. Appl. \textbf{45}, 930--953 (2024).

\bibitem{JMLR:v21:19-027}
Miolane, N., Guigui, N., Le Brigant, A., Mathe, J., Hou, B., Thanwerdas, Y., Heyder, S., Peltre, O., Koep, N., Zaatiti, H., Hajri, H., Cabanes, Y., Gerald, T., Chauchat, P., Shewmake, C., Brooks, D., Kainz, B., Donnat, C., Holmes, S., Pennec, X.: 
Geomstats: A Python Package for Riemannian Geometry in Machine Learning. 
\textit{Journal of Machine Learning Research} \textbf{21}(223), 1--9 (2020). 

\bibitem{schaefer2018local}
Schaefer, Alexander and Kong, Ru and Gordon, Evan M and Laumann, Timothy O and Zuo, Xi-Nian and Holmes, Avram J and Eickhoff, Simon B and Yeo, BT Thomas: Local-global parcellation of the human cerebral cortex from intrinsic functional connectivity MRI. Cerebral cortex \textbf{28}(9), 3095--3114 (2018)

\bibitem{VANESSEN20122222}
D.C. {Van Essen} and K. Ugurbil and E. Auerbach and D. Barch and T.E.J. Behrens and R. Bucholz and A. Chang and L. Chen and others: The Human Connectome Project: A data acquisition perspective. NeuroImage \textbf{62}(4), 2222-2231 (2012)

\bibitem{HCP:minimali}
Glasser, M.F. and Sotiropoulos, S.N. and Wilson, J.A. and Coalson, T.S. and Fischl, B. and Andersson, J.L. and Xu, J. and Jbabdi, S. and Webster, M. and Polimeni, J.R. and others: The minimal preprocessing pipelines for the Human Connectome Project. NeuroImage \textbf{80}, 105--124 (2013)

\bibitem{HumanConnectome}
Sporns, O., Tononi, G., and Kötter, R.: The human connectome: a structural description of the human brain. PLoS computational biology \textbf{1}(4), (2005)


\bibitem{DynamicConnectomes}
William Hedley Thompson, Per Brantefors, Peter Fransson: From static to temporal network theory: Applications to functional brain connectivity. Network Neuroscience \textbf{1}(2), 69-99 (2017)

\bibitem{pennec2006riemannian}
Pennec, X., Fillard, P., Ayache, N.:
A Riemannian framework for tensor computing. 
\textit{International Journal of Computer Vision} \textbf{66}, 41--66 (2006).
Springer.

\bibitem{arsigny2006log}
Arsigny, V., Fillard, P., Pennec, X., Ayache, N.:
Log-Euclidean metrics for fast and simple calculus on diffusion tensors. 
\textit{Magnetic Resonance in Medicine} \textbf{56}(2), 411--421 (2006).
Wiley Online Library.

\bibitem{johnson009}
C.~R.~Johnson and R.~Reams,
Scaling of symmetric matrices by positive diagonal congruence,
\emph{Linear and Multilinear Algebra},
vol.\ 57, no.\ 2, pp.\ 123–140, 2009.
Taylor \& Francis. \href{https://doi.org/10.1080/03081080600872327}{doi:10.1080/03081080600872327}.

\bibitem{marshall1968}
A.~W.~Marshall and I.~Olkin,
Scaling of matrices to achieve specified row and column sums,
\emph{Numerische Mathematik},
vol.\ 12, pp.\ 83–90, 1968.
\href{https://api.semanticscholar.org/CorpusID:121145311}{URL}.


\end{thebibliography}
%

\end{document}